\documentclass{article}
\usepackage{arxiv}
\usepackage{dcolumn}
\usepackage{mathptmx} 
\usepackage{multirow}
\usepackage{} 

\usepackage[utf8]{inputenc} % allow utf-8 input
\usepackage[T1]{fontenc}    % use 8-bit T1 fonts
\usepackage{hyperref}       % hyperlinks
\usepackage{url}            % simple URL typesetting
\usepackage{booktabs}       % professional-quality tables
\usepackage{amsfonts}       % blackboard math symbols
\usepackage{nicefrac}       % compact symbols for 1/2, etc.
\usepackage{microtype}      % microtypography
\usepackage{lipsum}
\usepackage{amssymb,amsmath,epsfig}
\newcommand{\beq}{\begin{equation}}
\newcommand{\eeq}{\end{equation}}
\newcommand{\bea}{\begin{eqnarray}}
\newcommand{\eea}{\end{eqnarray}}
\begin{document}
\begin{flushright}
Prepared for Physics of the Dark Universe
\end{flushright}
\title{Particle collisions in ergoregion of braneworld Kerr black hole}
\author{{Saeed Ullah Khan$^1$\thanks{saeedkhan.u@gmail.com}},\, {Misbah Shahzadi$^{2}$\thanks{misbahshahzadi51@gmail.com}}\, and {Jingli Ren$^{1}$\thanks{renjl@zzu.edu.cn}}\\
\normalsize$^1$School of Mathematics and Statistics, Zhengzhou University,\\\normalsize
Zhengzhou 450001, China.\\
\normalsize$^2$Department of Mathematics, COMSATS University Islamabad,\\\normalsize
Lahore Campus-54000, Pakistan.}

\author{{Saeed Ullah Khan$^1$\thanks{saeedkhan.u@gmail.com}} \,{Misbah Shahzadi$^{2}$\thanks{misbahshahzadi51@gmail.com}}\, and \, {Jingli Ren$^{1}$\thanks{renjl@zzu.edu.cn}}\vspace{0.2cm} \\\vspace{0.08cm}
$^1$School of Mathematics and Statistics, Zhengzhou University, Zhengzhou 450001, China.\\
$^2$Department of Mathematics, COMSATS University Islamabad, Lahore Campus-54000, Pakistan.}
\date{}
\maketitle
%%-------------------------------------------------%%
\begin{abstract}
This paper explores the neutral particle motion and collisional Penrose process in ergoregion of the braneworld Kerr black 
hole. We analyze the properties of event horizon, ergosphere and static limit. The particle collision in ergoregion via the Penrose process is investigated. Furthermore, we study the negative energy states and show that the sign of particle energy can be uniquely determined by the sign of angular momentum. In addition, we study the Wald inequality to determine the limits of energy extraction via the 
Penrose process and also find lower bound of the irreducible mass. The expression for the efficiency of energy extraction from the brane Kerr black hole is found. Finally, we compare our results with that obtained from the Kerr black hole. It is concluded that efficiency increases with the increase of rotation as well as brane parameter $b$ of the black hole.
\end{abstract}
%%---------------%%
%\textbf{PACS:} 04.70.-s; 52.30.Cv; 52.25.Xz.
\tableofcontents
%\textbf{PACS:} 04.70.-s; 97.60.Lf; 04.70.Bw.
%%----------------------------------------------------------%%
\section{Introduction}\label{Sec1}
%%---------------------------------------------------------%%
In recent years, researchers have made great contributions to the higher-dimensional string as well as M-theories, which are among the 
most aspiring approaches to the higher dimensional gravity theories \cite{Horava1,Horava2}. These theories describe gravity as a truly 
higher dimensional interaction that becomes effectively 4D at low enough energies. The braneworld models have been inspired by these 
theories, where the observable Universe is a 3-brane, on which the standard-model (non-gravitational) matter fields are confined, while 
the gravity fields enter into the extra spatial dimensions. The extra dimensions could modify the properties of a black hole (BH) 
and may have much larger size than that of the Planck length scale ($l_p \sim 10^{-33}$cm) \cite{Arkani-Hamed}. It is possible that 
these extra dimensions could have infinite size, just like in case of the braneworld model of Randall and Sundrum comprise of one extra 
spatial dimension \cite{Randall}. Consequently, these models may 
provide an effective solution to the hierarchy problems of the 
electroweak and quantum gravity scales, as these scales could become 
to be of the same orders ($\sim$ TeV) due to the large scale extra 
dimensions. Hence, the braneworld models can be tested by future 
collider experiments quite well, inclusive of the hypothetical mini 
BH construction of the TeV- energy scales \cite{Dimopoulos, 
Emparan}. In addition, the angular momentum of a BH in the vicinity 
of 4D general relativity (GR) is limited by the Kerr bound but in 
string theory, the required bound could split and compact objects 
like BH can spin faster \cite{Gimon}.

In the last few decades, a number of scientific research has been 
carried out to study the dynamics around Reissner-Nordström (RN), 
Kerr-Newman (KN) and modified BHs \cite{Z. Stuchlik4, Z. Stuchlik5, 
Sharif1, Sharif2, Oteeva}, which one can directly apply to the 
braneworld models, considering the effects of only positive tidal 
charge. Dadhich et al. \cite{Dadhich1} deduced that the RN spacetime 
is the exact solution of effective Einstein equations on brane, re-
defined as a BH with the effects of tidal charge rather than the electric charge. Kotrlová et al. \cite{Kotrlov} by examining the 
braneworld model, inferred that mass of the neutron star reduces due 
to the existence of negative tidal charge. Pugliese et al. \cite{Pugliese1,Pugliese2} studied the RN spacetime and extrapolates 
the existence of circular orbit with vanishing angular momentum. Exploring the circular geodesics in Kerr-Newman BH, it is concluded 
that BHs and naked singularities can be altered using the structure 
of stability’s regions \cite{Pugliese3}. The braneworld models were also studied by many researchers considering both negative as well 
as positive tidal charge effects \cite{Z. Stu6, Casadio, Blaschke, Z.Stuchlik3}. Stuchl{\'i}k and Hled{\'i}k \cite{Stuchlik3a} studied 
the properties of RN BH and naked singularity with non-zero 
cosmological constant and determined the photon escape cones. Grib and Pavlov \cite{Grib A2} explored the energy bounds within 
ergosphere of a BH and found that there is no extreme BH having a 
critical value of intrinsic angular momentum of the BH rotation. 
Nakao et al. \cite{Nakao} concluded that the Kerr superspinars can be stable against linear perturbations whereas, BHs on the Randall-
Sundrum brane could be stable against each type of perturbations \cite{To-st}. Stuchlik and Kolos \cite{Stuch-kolo} studied the 
chaotic scattering around a BH immersed in uniform magnetic field and observed that the strong acceleration of ionized particles 
towards ultra-relativistic velocities could be preferred in the 
direction nearby the magnetic field lines.

It is widely known that similar to the astrophysical conditions, the 
BH electric charge becomes negligible or vanished on small intervals 
of time due to its neutralization of accreting preferentially 
oppositely charged particles from an ionized matter of the accretion 
disc \cite{dovich, Misner}. This statement is also true in case of 
the braneworld models, that's why it is sufficient to examine the 
properties of brane Kerr BHs enriched with its tidal charge effects 
only. The consequences of tidal charge effects were also studied by 
some researchers in recent years for optical lensing in the weak 
field limits \cite{Keeton} and in the time delay effects 
\cite{B.ohmer}. It is observed that for the brane Kerr BH with fixed 
rotational parameter, the increasing values of negative tidal charge 
strengthens the relativistic effects \cite{Schee1}. Schee and 
Stuchl{\'i}k \cite{Schee2} deduced that the profiled lines in the 
framework of braneworld Kerr spacetime depend on spin as well as 
brane parameter and become wider by lowering the negative tidal charge. 

The process of energy extraction from rotating BHs is among the 
significant and aspiring problems in the fields of GR as well as in 
astrophysics. Penrose \cite{R.Penrose} introduce a truly accurate 
mechanism to extract energy from a rotating BH and related to the 
existence of the negative energy in ergoregion. It is found that 
energy extraction could be greater in case of the higher dimensional BHs as compared to the energy extortion from 4D Kerr BH 
\cite{Nozawa}. It is deduced that more energy can be extracted with rotating particle as compared to the non-rotating case 
\cite{Mukherjee}.

The Efficiency of energy extraction from a rotating BH via the 
Penrose process could be explained as (gain in energy)/(input 
energy). Efficiency of the Penrose process gets minimized around KN 
BH due to the vicinity of charge in comparison with the maximum 
efficiency limit of $20.7\%$ for that of the Kerr BH \cite{Bhat}. 
Parthasarathy et al. \cite{Parthasarathy} by studying rotating BH 
under the effect of magnetic field, extrapolate that, efficiency of 
the Penrose process could reach up to $100\%$, if an incoming 
particle splits near the static limit. Liu et al. \cite{Liu} found that in the case of non-Kerr BH deformation parameter increases the 
efficiency of energy extraction. Investigation of the Penrose 
mechanism around a regular rotating BH shows decreasing behavior in 
the efficiency of the energy extraction for increasing values of the 
electric charge \cite{Toshmatov}. Liu and Liu \cite{Liu2} studied 
the Penrose mechanism with rotating particles and deduced that 
efficiency of the energy extraction monotonically increases as the 
particle rotation increases. Dadhich et al. \cite{Dadhich} explored 
the Penrose process in the presence of magnetic field and concluded 
that efficiency of the process increases with the magnetic charge. Shahzadi et al. \cite{M.Shahzadi} explored the particle motion near 
Kerr-MOG (modified theory of gravity) BH and found that the efficiency of energy extraction can be enhanced with the increase of 
the dimensionless parameter of the theory.

In this article, we investigate the neutral particle motion and the collisional Penrose process within ergoregion of the braneworld Kerr BH. The article is organized as follows: the coming section will briefly summarize the braneworld Kerr BH, its properties and effects of the tidal charge on ergoregion, static limit and on horizons of the BH. In section \ref{Sec3}, we derive the equations of motion and discuss the angular velocity of particles in the vicinity of ergoregion. In the next section, we study the collisional Penrose process and investigate the negative energy states, Wald inequality 
as well as the efficiency of energy extraction. Finally, we summarize our results in the last section.
%%---------------------------------------------------------%%
\section{Braneworld Kerr Black Hole}\label{Sec2}
%%---------------------------------------------------------%%
The braneworld kerr BH is an axially symmetric, stationary and 
asymptotically flat solution of the effective Einstein equations on the brane. The spacetime geometry of the braneworld Kerr BH could be 
outlined by the metric with Boyer-Lindquist coordinates as \cite{Aliev}
\begin{equation}
ds^2 = g_{tt}dt^{2} + 2 g_{t\phi}dtd\phi + g_{rr}dr^{2} + g_{\theta\theta}d\theta^2 + g_{\phi\phi}d\phi^2,\label{1}
\end{equation}
with
\begin{eqnarray}
g_{tt}&=&-\left(\frac{\Delta-a^{2}\sin^{2}\theta}{\Sigma}\right),
\quad g_{rr}=\frac{\Sigma}{\Delta}, \quad
g_{\theta\theta}=\Sigma, \\\nonumber
g_{\phi\phi}&=&\frac{\sin^{2}\theta}{\Sigma}\left[(r^{2}+a^{2})^{2}-\Delta
a^{2}\sin^{2}\theta\right], \\\nonumber
g_{t\phi}&=&\frac{a\sin^{2}\theta}{\Sigma}\left[\Delta-(r^{2}+a^{2})\right],
\end{eqnarray}\label{2}
where
\begin{eqnarray}
\Delta = r^{2} - 2Mr + a^{2} + b,\quad
\Sigma = r^{2}+a^{2}\cos^{2}\theta.
\end{eqnarray}\label{3}
Here, $M$ and $a$ represents the mass and spin parameter of the BH, 
respectively. In the case of braneworld Kerr BH, the impact of tidal effects from the bulk is expressed by a single parameter known as 
the tidal charge $b$, which can take both positive as well as negative values. It can be observed that metric \eqref{1} exactly 
looks like the KN BH with the replacement of $b$ by $Q^2$ 
(representing electric charge of the KN BH). Moreover, it is 
worthwhile to note that for different values of the parameter $b$, Eq. \eqref{1} assumes special cases, i.e., it reduced to the Kerr BH 
if $b=0$; for $b>0$, we obtained the KN BH; and finally for $b<0$, it reduced into the non-standard KN BH with negative tidal effects. 
Furthermore, for both $a$, $b=0$, the metric \eqref{1} reduced to the Schwarzschild BH.
%%---------------------------------------------------------%%
\subsection{Ergoregion, Horizon and Static Limit}
%%---------------------------------------------------------%%
The ergoregion (section lies between the horizons and the static limit of a BH) play an important role in astrophysics, as the 
Hawking radiation could be examined in this region. Moreover, the ergoregion of a BH is also momentous because of the Penrose process, 
as it takes place in this region. The static limit surface ($r_{es}$) is also termed as the infinite redshift surface, in which the 
time-translation killing vector becomes null. When the time-like 
geodesics crossed the static limit surface they changed into the 
space-like geodesics. On substituting $g_{tt}=0$, the ergosphere 
turns out to be
\begin{equation}\nonumber
r_{es} = M \pm \sqrt{M^2 - a^2 \cos^2\theta - b}.
\end{equation}
The existence of static limit requires that
\begin{equation}\nonumber
M^2 \geq a^2 \cos^2{\theta}+b.
\end{equation}
The horizons of Eq. \eqref{1} can be found by solving $\Delta=0$ as
\begin{equation}\nonumber
r_{\pm} = M\pm \sqrt{M^2-a^2-b}.
\end{equation}
The event horizon will exist if
\begin{equation}\nonumber
M^{2} \geq a^{2}+b.
\end{equation}
%%----------------------------------------------------------%%
\begin{figure*}
\vspace{-1.0cm}
\includegraphics[width=\hsize]{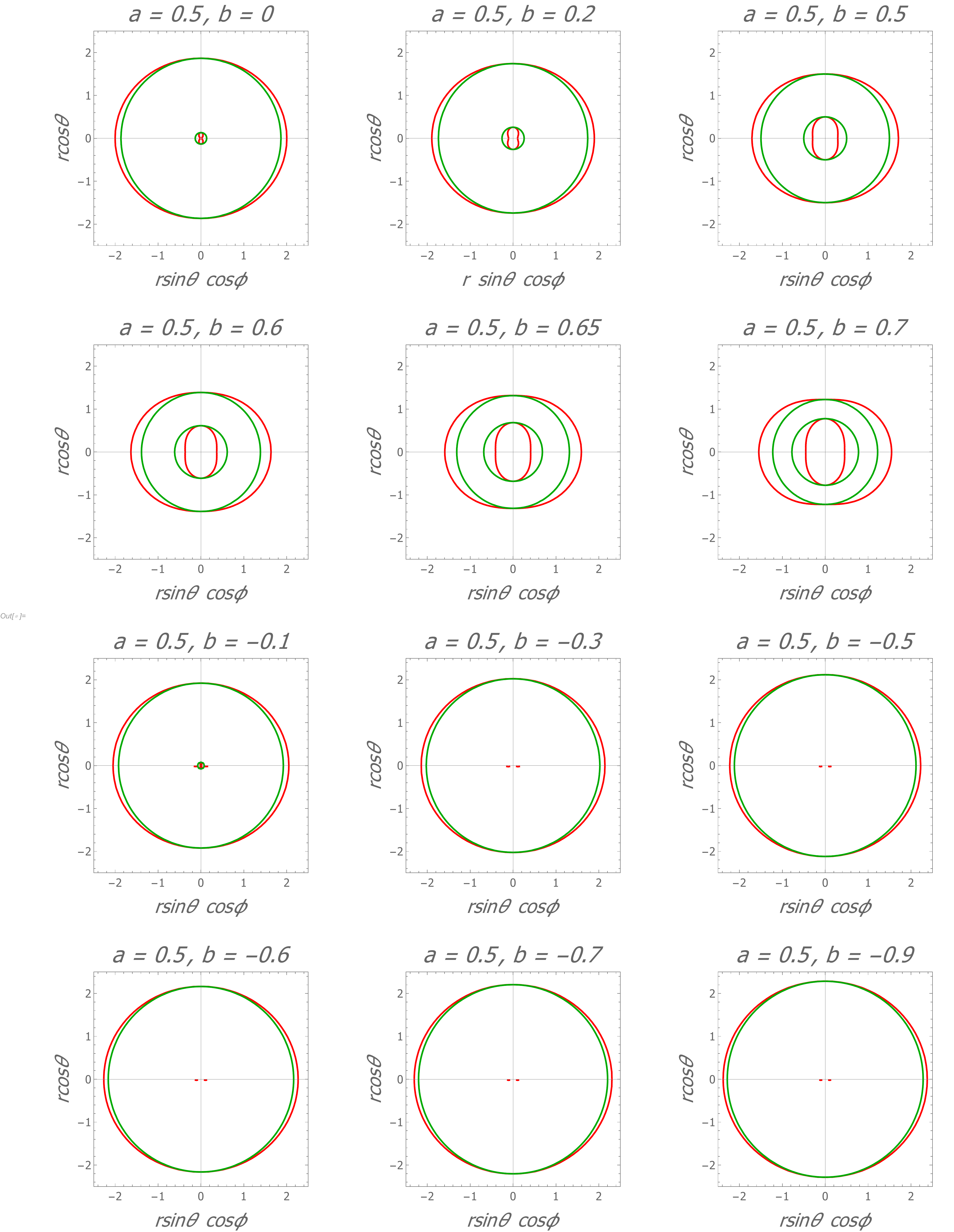}
\caption{The ergosphere and horizons in the xz-plane for different values of tidal charge $b$.} \label{ergo1}
\end{figure*}
%%------------------------------------------------------------%%
\begin{figure*}
\includegraphics[width=\hsize]{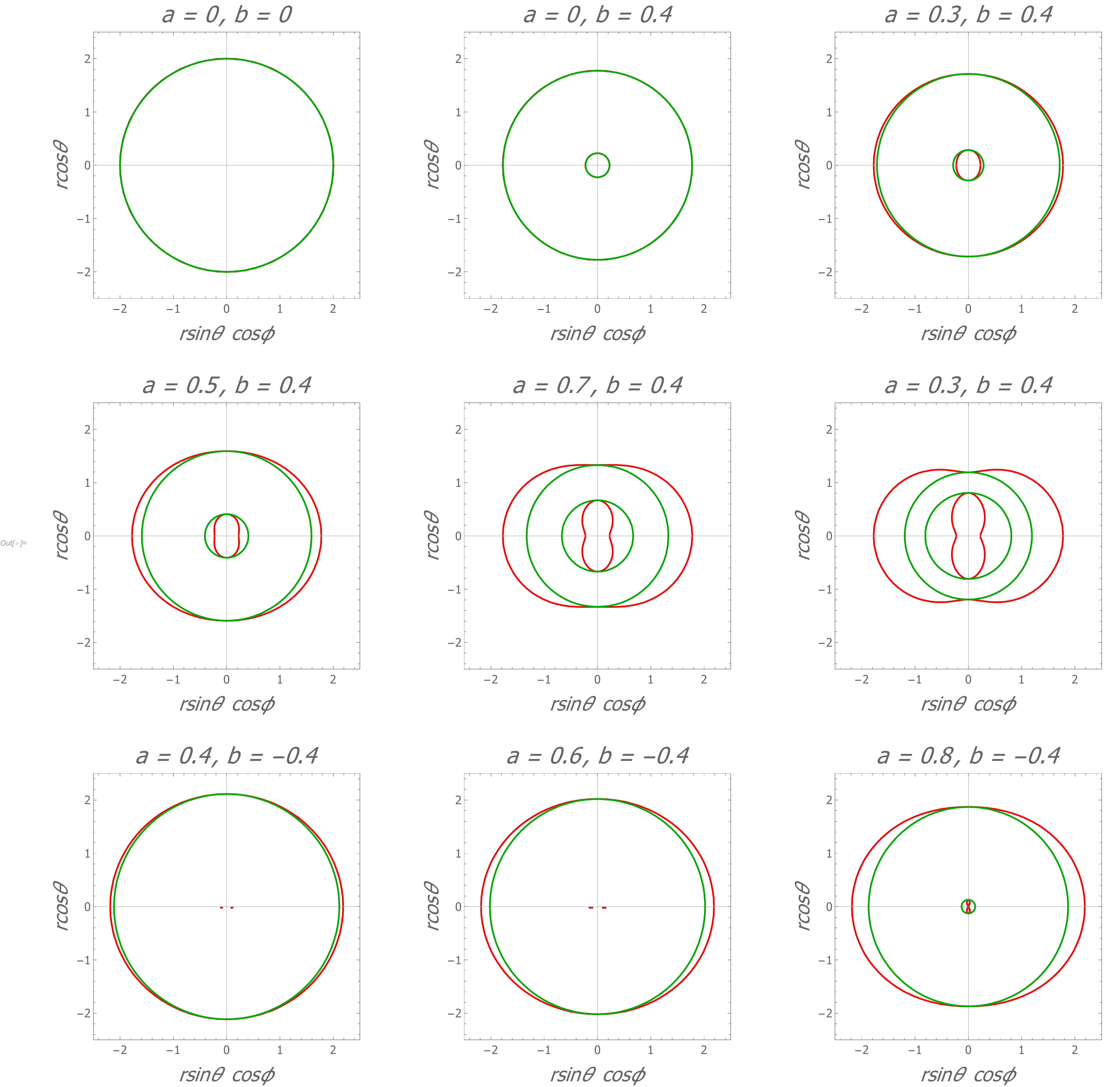}
\caption{The ergosphere and horizons in the xz-plane for different values of spin parameter $a$.}\label{ergo2}
\end{figure*}
%%------------------------------------------------------------%%
The effects of positive tidal charge tend to reduce gravitational 
field and in this case, similar horizons structure can be found as 
in the KN BH. However, the negative tidal charge may lead us to some 
new interesting features of the BH. For $a\rightarrow M$ and $b<0$, the horizon radius $r_+ \rightarrow (M+\sqrt{-b})>M$, and this 
condition does not hold in GR. The extremal condition can be found 
at $M^{2} = a^{2}+b$ and in case of the negative tidal charge, the 
extreme horizon $(r_+=M)$ correspond to the BH with $b=-M^2$ and $a=\sqrt{2} M$ can be obtained. Therefore, bulk effects on the brane 
could provide the mechanism for spinning up the BH on the brane, consequently, its spin parameter exceeds its mass and this type of 
situation is impossible in GR.

The numerical values of event horizon, static limit and ergoregion 
of the braneworld Kerr BH at different values of $b$ and $a$, are 
given in Table. {\bf1}, and the corresponding structures are 
depicted in Figs. {\bf\ref{ergo1}} and {\bf\ref{ergo2}}. It is 
observed that the sign of tidal charge has significant effects on the shape of ergoregion as well as on the horizons. Figure {\bf
\ref{ergo1}} shows that for positive values of the tidal charge the ergoregion becomes thick, while both of its radii for the static 
limit surface and event horizon decreases. On the other hand, for 
negative values of the tidal charge, the ergoregion decreases whereas the horizons increases. In Fig. {\bf\ref{ergo2}}, it is 
demonstrated that the area of ergoregion increases by increasing the 
values of both $a$ and $b$ and its radii get decreases. In addition, 
it is found that the braneworld Kerr BH has greater radii and 
thicker ergoregion as compared to the Kerr BH.
%%----------------------------------------------------------%%
\begin{table*}
\begin{center}
\textbf{Table 1:} The ergoregion ($\delta=r_{+es}-r_+$), event horizon ($r_+$) and static limit ($r_{+es}$).
\\
\small\addtolength{\tabcolsep}{0.0pt}
\scalebox{1}{
\begin{tabular}{|c| c| c| c| c|}
\hline \hline \noalign{\smallskip}
 &  {a=0.2}  &  {a=0.3}  &   {a=0.4} & {a=0.5}  \\[.6em]
\hline \noalign{\smallskip}
$b$ & $r_+ \quad \quad r_{+es} \quad \quad \delta$ & $r_+ \quad \quad r_{+es} \quad \quad \delta$ &$r_+ \quad \quad r_{+es} \quad \quad \delta$ &$r_+ \quad \quad r_{+es} \quad \quad \delta$   \\[.6em]
\hline
\\[-.9em]
-0.7 & 2.2884  2.2962  0.0078& 2.2689  2.2866 0.0177 &2.2410 2.2728  0.0318 &2.2042 2.2550  0.0508 \\[.7em]

-0.5 & 2.2083  2.2166  0.0083& 2.1874  2.2062 0.0188 &2.1576 2.1917  0.0341 &2.1180 2.1726  0.0550 \\[.7em]

-0.3 & 2.1225  2.1314  0.0089& 2.1000  2.1203 0.0203 &2.0677 2.1050  0.0368 &2.0247 2.0840  0.0593 \\[.7em]

-0.1 & 2.0296  2.0392  0.0097& 2.0050  2.0271 0.0221 &1.9695 2.0100  0.0404 &1.9220 1.9872  0.0652 \\[.7em]

0.0 &  1.9798  1.9900  0.0102& 1.9540  1.9772 0.0233 &1.9165 1.9592  0.0427 &1.8660 1.9354  0.0694 \\[.7em]

0.1 &  1.9274  1.9381  0.0107& 1.9000  1.9247 0.0247 &1.8602 1.9055  0.0453 &1.8062 1.8803  0.0741 \\[.7em]

0.3 &  1.8124  1.8247  0.0123& 1.7810  1.8093 0.0283 &1.7349 1.7874  0.0526 &1.6708 1.7583  0.0875 \\[.7em]

0.5 &  1.6782  1.6928  0.0146& 1.6403  1.6745 0.0342 &1.5831 1.6481  0.0650 &1.5000 1.6124  0.1124 \\[.7em]

0.7 &  1.5100  1.5292  0.0192& 1.4583  1.5050 0.0467 &1.3742 1.4690  0.0948 &1.2236 1.4183  0.1947 \\[.5em]
\hline \hline
\end{tabular}}
\end{center}\label{Tab1}
\end{table*}
%%%---------------------------------------------------------%%
\section{Particle Dynamics}\label{Sec3}
%%%---------------------------------------------------------%%
We assume the neutral particle motion in background of the braneworld Kerr BH and limited our investigation to the case of 
orbits situated on an equatorial plane. The governing equation of 
geodesics can be acquired using the Lagrangian equation as
\begin{equation}
\mathcal{L} = \frac{1}{2}g_{\mu\eta}\dot{x}^{\mu}\dot{x}^{\eta},
\end{equation}
where dot means $\partial / \partial \tau$ ($\tau$ is the proper 
time) and $\dot{x}^{\mu}$ represents the four-velocity. The generalized momenta for (\ref{1}) can be written as
\begin{eqnarray}\label{6}
-p_{t}&=&g_{tt}\dot{t}+g_{t\phi}\dot{\phi}=E,\\\label{7}
p_{\phi}&=&g_{t\phi}\dot{t}+g_{\phi\phi}\dot{\phi}=L,
\\\nonumber p_{r}&=&g_{rr}\dot{r},
\end{eqnarray}
where $E$ and $L$ are interpreted as the energy and angular momentum of a particle associated with the Killing vector fields
$\xi_{t}=\partial_{t}$ and $\xi_{\phi}=\partial_{\phi}$,
respectively. Since the Lagrangian is independent of the coordinates
$t$ and $\phi$, so $p_{t}$ and $p_{\phi}$ are conserved along
geodesics and hence specifies the stationary as well as axisymmetric
properties of the Kerr braneworld BH. From Eqs. (\ref{6}) and (\ref{7}), we
obtain
\begin{eqnarray}\label{energy}
\dot{t}&=&\frac{1}{r^2 \Delta}\left[E a^2 (b-r (2 M + r)) + a L (b-2 M r) - E r^4\right],\\\label{angmomentum}
\dot{\phi}&=&\frac{1}{r^2 \Delta}\left[E a (b - 2 M r) + L (b+r (r-2 M))\right].
\end{eqnarray}
The Hamiltonian for neutral particle motion can be written as
\begin{equation}\nonumber
H=p_{t}\dot{t}+p_{r}\dot{r}+p_{\phi}\dot{\phi}-\mathcal{L}.
\end{equation}
For Eq. (\ref{1}), the above equation turns out to be
\begin{eqnarray}\nonumber
2H&=&-\left(g_{tt}\dot{t}+g_{t\phi}\dot{\phi}\right)\dot{t}+
\left(g_{t\phi}\dot{t}+g_{\phi\phi}\dot{\phi}\right)\dot{\phi}+g_{rr}\dot{r}^{2}\\\label{hamiltonian}
&=&E\dot{t}+L\dot{\phi}+\frac{r^{2}}{\Delta}\dot{r}^{2}=\epsilon=constant,
\end{eqnarray}
where $\epsilon=-1, 0, 1$ specify the timelike,  null (lightlike) and
spacelike geodesics. Substituting Eqs. (\ref{energy}) and (\ref{angmomentum}) into
(\ref{hamiltonian}), we find the radial equation of motion
\begin{eqnarray} \label{radial}
\dot{r}^{2}=E^{2}+\frac{1}{r^{4}}(2Mr-b)(aE-L)^{2}+\frac{1}{r^{2}}(a^{2}E^{2}-L^{2})+\epsilon
\frac{\Delta}{r^{2}}.
\end{eqnarray}
Eqs. (\ref{energy})-(\ref{radial}) are very important as they can be 
used to discuss different properties associated with the particle 
motion near this BH.
%%%--------------------------------------------------------%%
\subsection{Angular Velocity of a Particle in Ergoshpere}
%%%--------------------------------------------------------%%
In this section, we are interested to investigate particle angular 
velocity ($\Omega = d\phi/dt$) and its limitations within the 
premises of ergosphere with the condition of $ds^2 \geq 0$. Hence
\begin{equation}\label{Ang1}
g_{tt}dt^2+g_{t\phi }dt d\phi+g_{\phi \phi}d\phi^2 \geq 0,
\end{equation}
and the angular velocity must satisfy the constraints $\Omega_+\leq\Omega\leq \Omega_-$ \cite{Lightman}, where
\begin{equation}\label{Ang2}
\Omega_{\pm}=\frac{ -g_{t\phi}\pm\sqrt{g_{t\phi}^2-g_{tt}g_{\phi\phi}}}{g_{\phi \phi}}.
\end{equation}
On the boundary of ergoshpere, $g_{tt}=0$ and $\Omega_+$ =0, while 
inside the ergoshpere, $g_{tt}<0$ and $\Omega_{\pm}>0$, and every 
particle move in the direction of BH rotation \cite{Misner, 
Chandrasekhar}. For the brane Kerr BH, on substituting the values of $g_{tt}$, $g_{t\phi}$ and $g_{\phi \phi}$ form Eq. \eqref{2} into  
Eq. \eqref{Ang2}, we have
\begin{equation}\label{Ang3}
\Omega_{\pm}=\frac{a \sin{\theta}(b-2Mr)\pm\Sigma\sqrt{\Delta}}{\sin{\theta}[a^2 \sin^2{\theta}\Delta-(r^2+a^2)^2]}= \omega \pm \frac{\Sigma\sqrt{\Delta}}{\sin\theta[a^2 \sin^2{\theta}\Delta-(r^2+a^2)^2]},
\end{equation}
where
\begin{equation*}
\omega= \frac{a(b-2Mr)}{a^2 \sin^2{\theta}\Delta-(r^2+a^2)^2}.
\end{equation*}
Approaching the event horizon, we obtain
%%%---------------------------------------------------------%%
\begin{figure*}
    \begin{minipage}[b]{0.58\textwidth} \hspace{0.0cm}
        \includegraphics[width=0.8\textwidth]{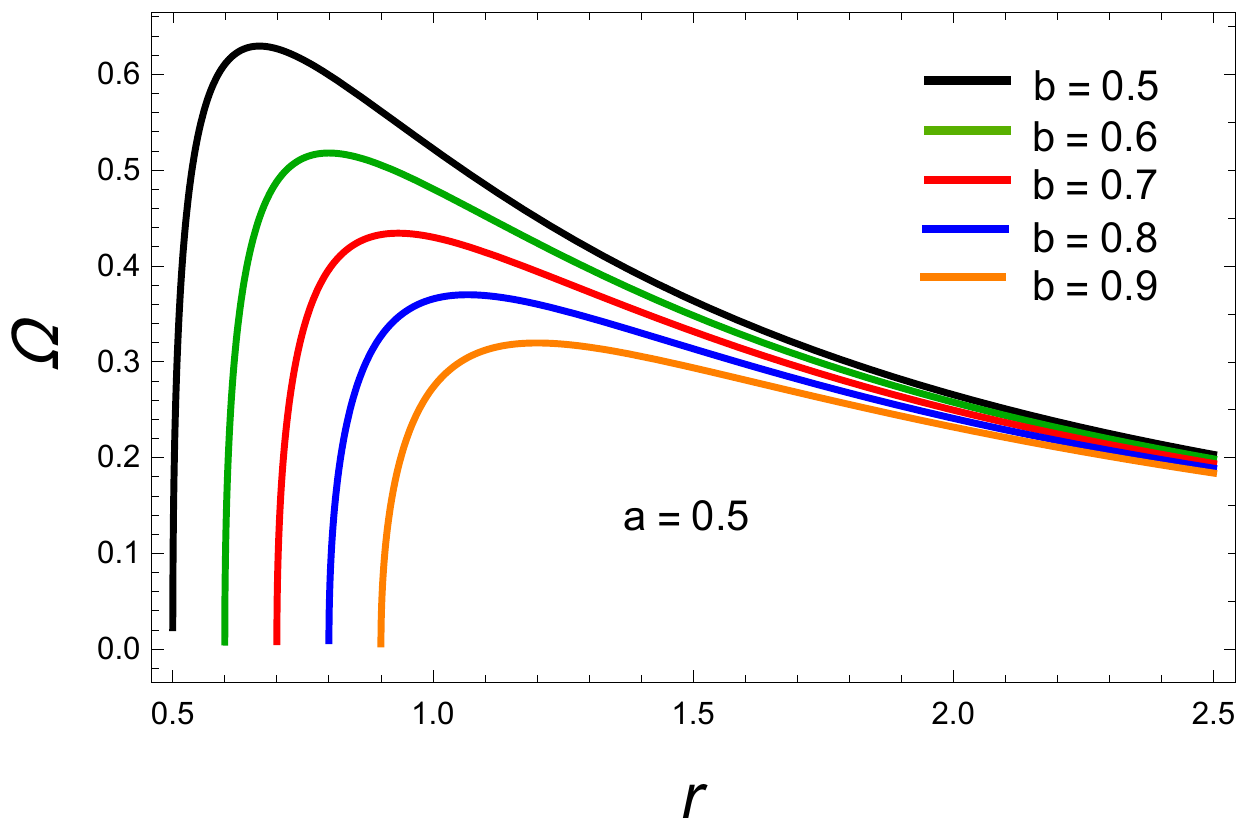}
    \end{minipage}
    \vspace{0.35cm}
        \begin{minipage}[b]{0.58\textwidth} \hspace{-1.5cm}
       \includegraphics[width=.8\textwidth]{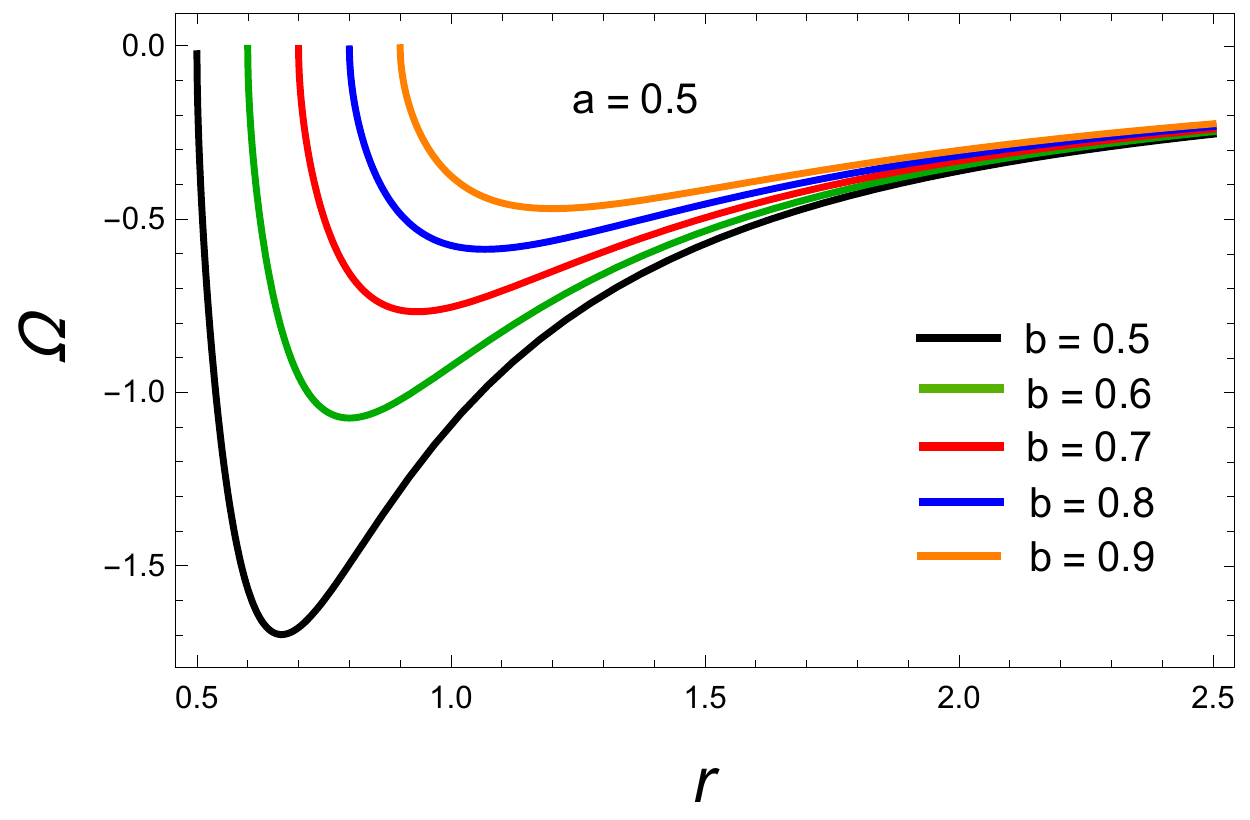}
    \end{minipage}
%%------------------------%%
\begin{minipage}[b]{0.58\textwidth} \hspace{-0.0cm}
        \includegraphics[width=0.8\textwidth]{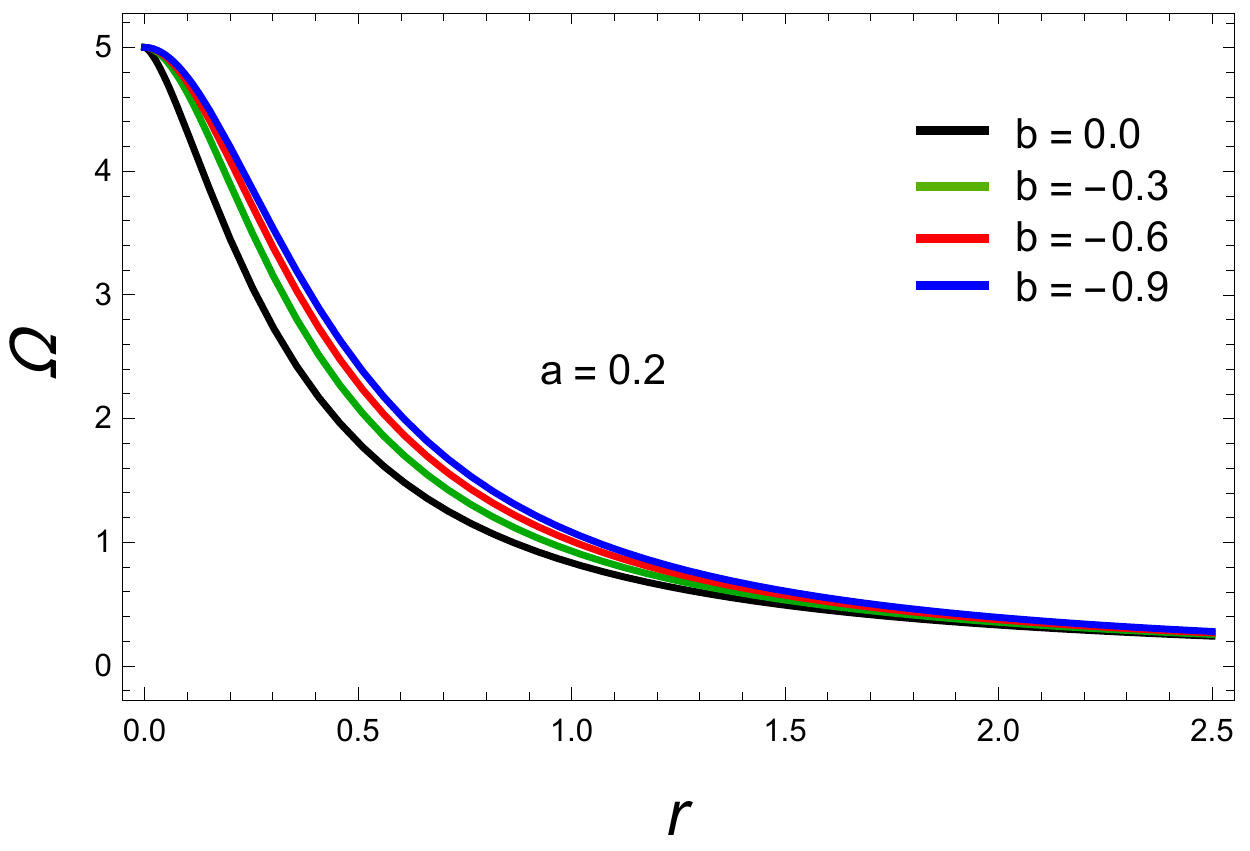}
    \end{minipage}
        \begin{minipage}[b]{0.58\textwidth} \hspace{-1.5cm}
       \includegraphics[width=.8\textwidth]{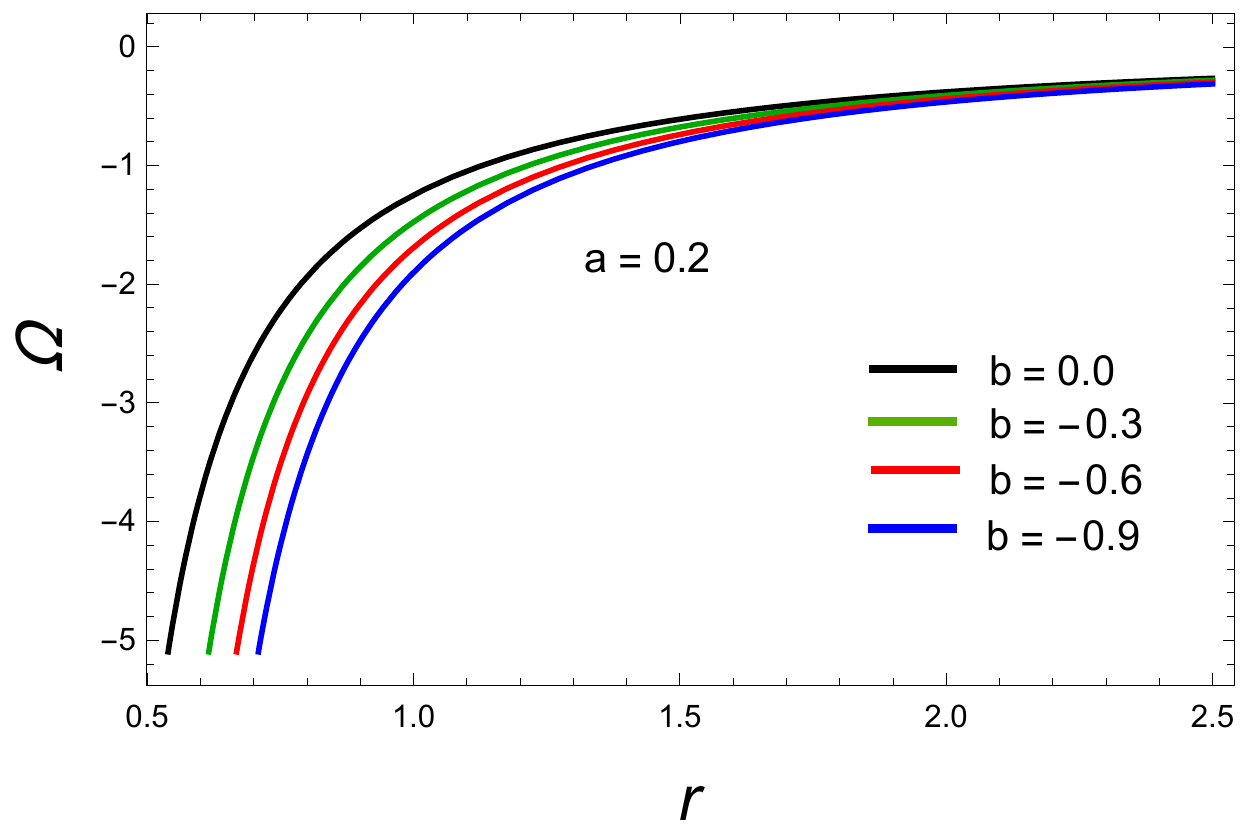}
    \end{minipage}
    \caption{The behavior of angular velocity $\Omega$ as a function of $b$ verses $r$.}\label{Ang01}
\end{figure*}
%%%----------------------------------------------------------%%
\begin{equation*}
\lim\limits_{r \to r_{+}} \Omega_{+} = \lim\limits_{r \to r_{+}} \Omega_{-}=\omega_{bh}=\frac{a(2Mr_{+}-b)}{r_{+}^4 + a^2 r_{+}^2 - a^2 (2Mr_{+} -b)}.
\end{equation*}
Here, $\omega_{bh}$ denotes angular velocity of the BH rotation and 
the angular velocity of particle turns out to be
\begin{equation}\label{Ang4}
\Omega=\frac{d\phi}{dt}=\frac{(r^2-2Mr+b)L+aE(2Mr-b)}{[r^2(r^2+a^2)+a^2(2Mr-b)]E-aL(2Mr-b)}.
\end{equation}
The specific energy, as well as the specific angular momentum of 
circular orbits at a given radius $r$, can be written as \cite{Stuchlikk}
\begin{eqnarray}\label{Ang5}
E&=&\frac{(r^2+b-2Mr\pm a\chi)}{r\left[r^2+2b-3Mr\pm 2a\chi \right]^{1/2}}, \\\label{Ang6}
L&=&\pm\frac{((r^2+a^2\mp 2a\chi)\mp ba)}{r\left[r^2+2b-3Mr\pm2a\chi \right]^{1/2}} ,
\end{eqnarray}
where $\chi=\sqrt{Mr-b}$, while the upper and lower signs correspond 
to the corotating and counter-rotating orbits, respectively. 
Substituting Eqs. \eqref{Ang5} and \eqref{Ang6} into \eqref{Ang4}, we obtain
\begin{equation}
\Omega=\frac{\mp\sqrt{Mr-b}}{(r^2\mp a\sqrt{Mr-b})}.
\end{equation}
The graphical behavior of angular velocity $\Omega$ is depicted in 
Fig. {\bf \ref{Ang01}}.  It is observed that angular velocity 
decreases as the brane parameter $b$ increases (both upper and lower 
row left panel), whereas increases as the brane parameter $b$ 
increases (both upper and lower row right panel).
%%----------------------------------------------------------%%
\section{Collisional Process in Ergoregion}\label{Sec4}
%%----------------------------------------------------------%%
The collisional process in rotating and charged BHs is an important 
and interesting issue of GR. During the collision process, energy 
can be extracted from a BH. There are a number of different 
techniques that can be used to explore the process of energy 
extraction from a rotating BH. Penrose process \cite{R.Penrose} is 
among the most rigorous and efficient energy extraction techniques 
than those of the nuclear reactions. In this mechanism, a particle 
enters into ergosphere having positive energy, subdivided into two 
particles, one of them follows the trajectory of negative energy 
whereas the other escape to infinity having more energy as compared 
to the incident ones. If the particles are involved in the Penrose process then the necessary and sufficient condition to extract 
energy from a BH is the absorption of particles with negative 
energies as well as angular momentum.
%%%---------------------------------------------------------%%
\subsection{Negative Energy States}
%%%---------------------------------------------------------%%
The negative energy states may occur due to the counter-rotating 
orbits (similar to that of the Kerr BH) as well as due to the 
electromagnetic interactions (similar to that of the RN BH) \cite{M.Shahzadi}. It is of great interest to find out the energy 
limits of a particle which they have at a particular location. The 
radial equation (\ref{radial}) could be rewritten as
%%----------------------------------------------------------%%
\begin{figure*} \vspace{-0.0cm}
\begin{minipage}[b]{0.58\textwidth} \hspace{-0.cm}
\includegraphics[width=.8\textwidth]{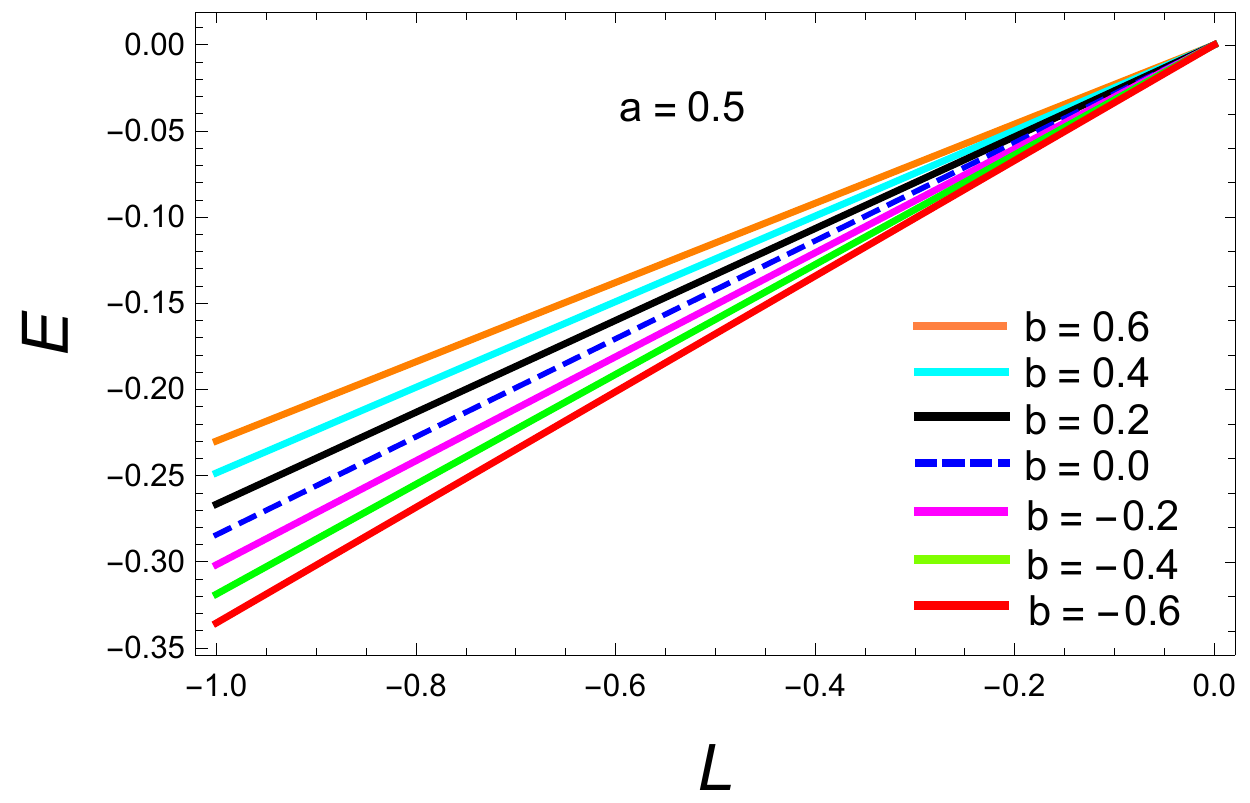}
\end{minipage}
\vspace{0.35cm}
\begin{minipage}[b]{0.58\textwidth} \hspace{-1.5cm}
\includegraphics[width=0.8\textwidth]{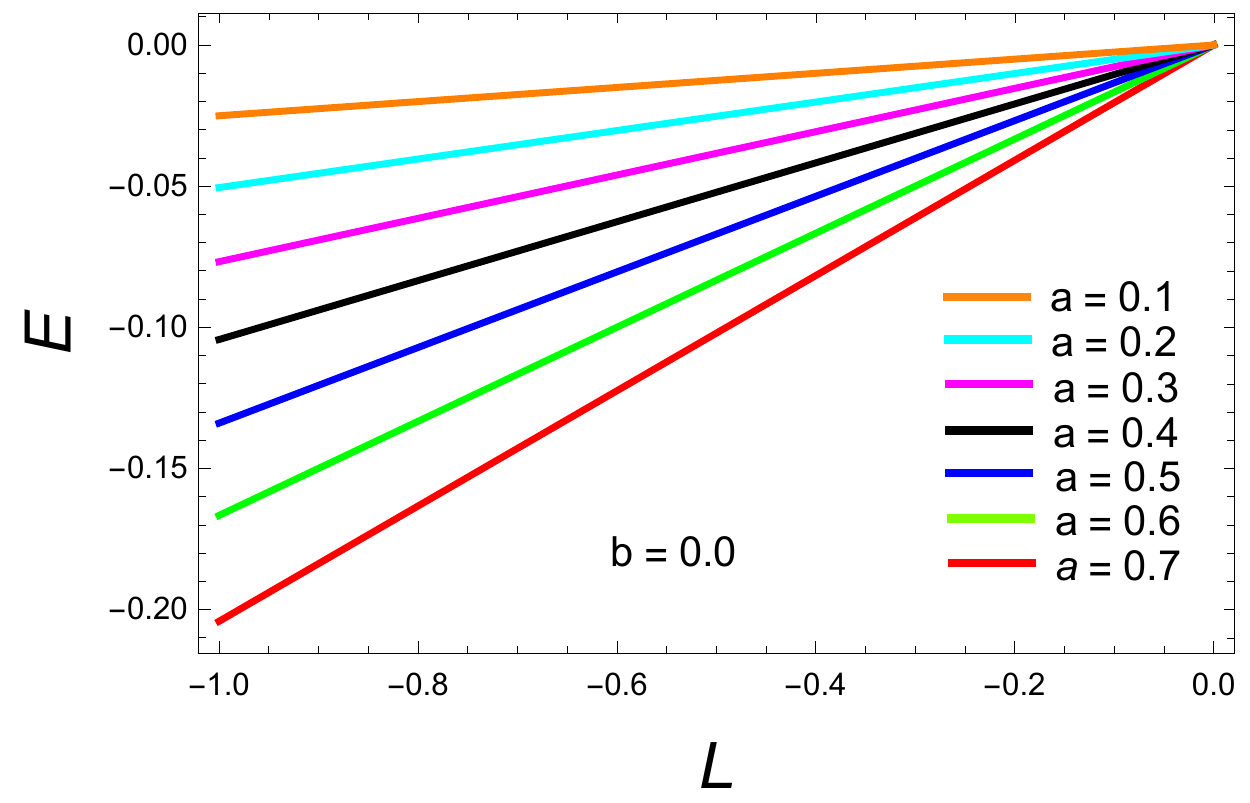}
\end{minipage}
%%-----------------------%%
\begin{minipage}[b]{0.58\textwidth} \hspace{-0.0cm}
\includegraphics[width=0.8\textwidth]{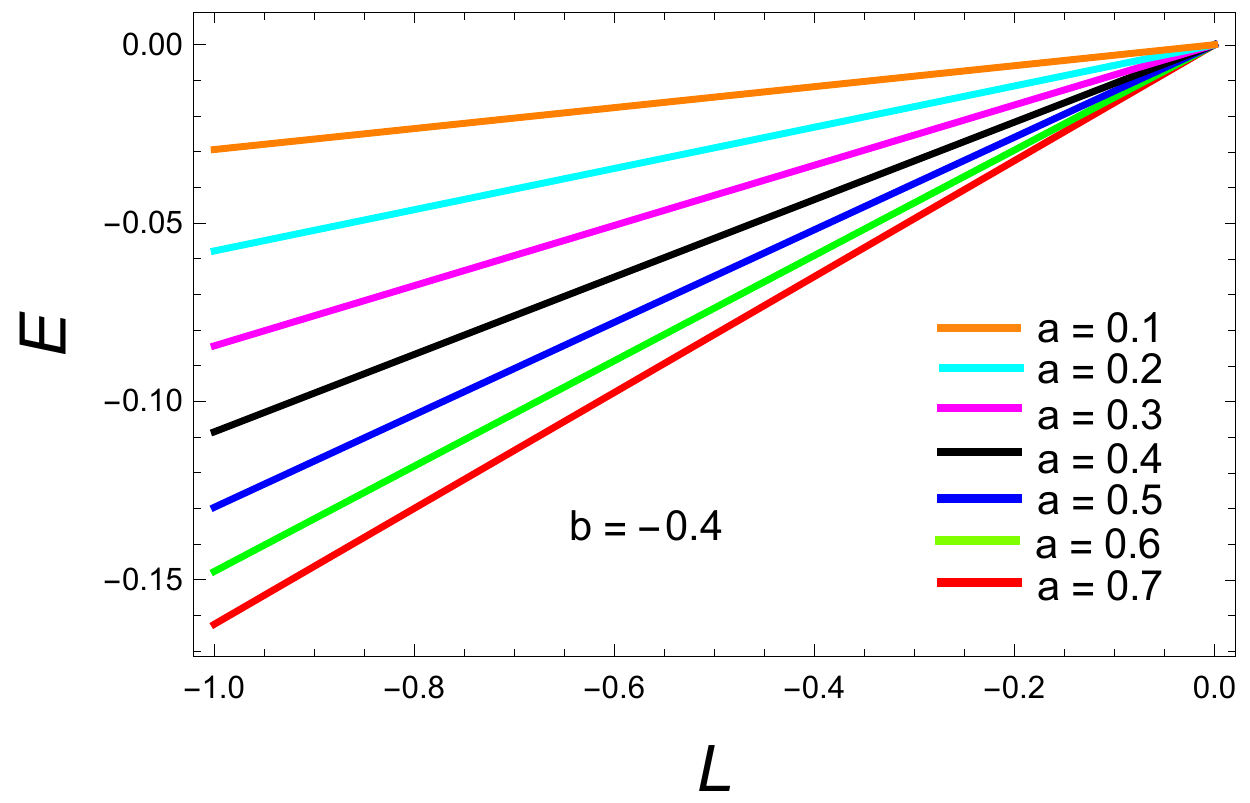}
\end{minipage}
\begin{minipage}[b]{0.58\textwidth} \hspace{-1.5cm}
\includegraphics[width=.8\textwidth]{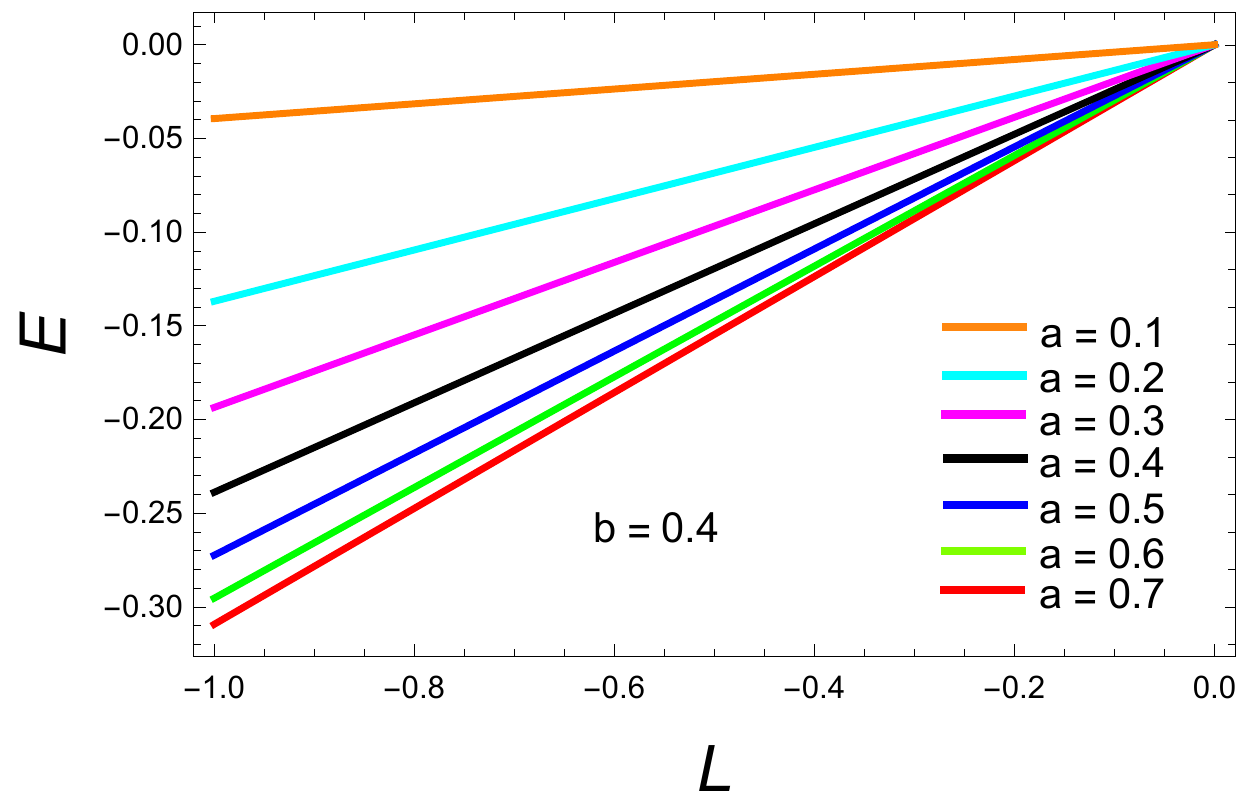}
\end{minipage}
\caption{Graphical representation of the negative energy states $E$ versus the angular momentum $L$.}\label{NE01}
\end{figure*}
%%----------------------------------------------------------%%
\begin{eqnarray}\nonumber
E^{2}[(r^{2}&+&a^{2})r^{2}+a^{2}(2 M r-b)] - 2 a E L(2Mr-b)
\\ &-&L^{2}(r^{2}-2 M r + b)+r^{2}\epsilon \Delta=0.\label{negenergy}
\end{eqnarray}
The values of $E$ and $L$ can be obtained from the above equation as
\begin{eqnarray} \label{NE0}
E&=&\frac{a L(2 M r-b) \pm \mathcal{Z}_{1}
\sqrt{\Delta}} {r^{4}+a^{2}(r^{2}+2Mr-b)},\label{NE1}
\\ L&=&\frac{-a E(2 M r-b)\pm \mathcal{Z}_{2}
\sqrt{\Delta}} {(r^{2}-2 M r+b)},
\end{eqnarray}
where
\begin{eqnarray} \nonumber
\mathcal{Z}_{1}&=&\sqrt{L^{2}r^{4}-[r^{4}+a^{2}(r^{2}+2Mr-b)]\epsilon r^{2}},\\\nonumber
\mathcal{Z}_{2}&=&\sqrt{r^{4}E^{2}+(r^{2}-2Mr+b)\epsilon
r^{2}}.
\end{eqnarray}
The following expression has been used to obtain the above results
\begin{eqnarray}\label{NE2}
r^{4}\Delta - a^{2}(2Mr-b)^{2}=(r^{4}+a^{2}(r^{2}+2Mr-b))(r^{2}-2Mr+b).
\end{eqnarray}
The conditions for negative energy can be deduced from Eq. (\ref{NE1}).
We set an energy $E=1$ and choose $+$ sign of Eq. (\ref{NE1}). For 
$E<0$, it also necessary that $L<0$, and
\begin{eqnarray}
a^{2}L^{2}\left(2 M r-b\right)^{2}>r^{2}\Delta[r^{2}L^{2}-\left(r^{4}+a^{2}(r^{2}+2Mr-b)\right)\epsilon].\label{NE3}
\end{eqnarray}
Using Eq. (\ref{NE2}), expression (\ref{NE3}) can be written as
\begin{eqnarray}
\left[r^{4}+a^{2}\left(r^{2}+2Mr-b\right)\right]\left[(r^{2}-2Mr
+b)L^{2}-\epsilon
r^{2}\Delta\right]<0.\label{NE4}
\end{eqnarray}
It can be concluded from inequality (\ref{NE4}) that $E<0 \Leftrightarrow L<0$,
and
\begin{equation}
\left(\frac{r^{2}-2Mr+b}{r^{2}}\right)<\frac{\Delta \epsilon}
{L^{2}}.
\end{equation}
For $b=0$, the above inequality reduced to the case of Kerr BH \cite{Chandrasekhar}. The behavior of negative energy $E$ versus 
angular momentum $L$ is illustrated in Fig. \textbf{\ref{NE01}}. The 
top left graph shows an increase in negative energy $E$ for 
decreasing values of $b$. The behavior of negative energy at $b=0$, $b=-0.4$ and $b=0.4$ are described in the top right, bottom left and 
bottom right panels respectively, at different values of the spin parameter $a$. Furthermore, it is concluded that the BH negative 
energy increases with the increase of its rotation and brane 
parameter $b$.

The irreducible mass of a BH is one of the consequences of energy 
extraction from the BH. When a particle of negative energy enters 
into the BH, the mass of the BH changes by a quantity $\delta M = E$ 
\cite{Abdujabbarov}. There is no upper bound on $\delta M$ as it can be increased by increasing mass of the injected particle. However, 
there is a lower bound on $\delta M$ and each incident particle 
having negative energy decreases the mass of the BH until its 
irreducible mass. The lower bound of $\delta M$ can be found using Eq. (\ref{NE1}). At horizon, the discriminant of Eq. (\ref{NE0}) is 
zero and thus we obtained the lower limit
\begin{eqnarray}
\delta M = \frac{a L(2 M r_{+}-b)} {r_{+}^{4}+a^{2}(r_{+}^{2}+2Mr_{+}-b)}\label{NE5}.
\end{eqnarray}
From Eq. (\ref{NE5}), it is concluded that to extract energy from a 
BH, the injected particle must have negative angular momentum and 
the brane parameter $b$ effects the value of $\delta M$.
%%%---------------------------------------------------------%%
\subsection{The Wald Inequality}
%%%---------------------------------------------------------%%
It is very important to study the limits of energy extraction by the 
Penrose process. Wald \cite{Wald} derived an inequality which can 
discuss the limitations of energy extraction by this process. To 
obtain these limits for the brane Kerr BH, consider a particle with 
specific energy $\mathcal{E}$ and four velocity $U^{\alpha}$, breaks 
up into fragments. Let $\mathcal{\varepsilon}$ be the specific 
energy and $u^{\alpha}$ be the four-velocity of one of the fragments. To derive the limits of $\mathcal{\varepsilon}$, consider 
an orthonormal tetrad frame $e_{j}^{\alpha}$, in which $U^{\alpha}$ 
coincides with $e_{0}^{\alpha}$ and the remaining basis vectors are $e_{(i)}^{\alpha} (i=1,2,3)$. In this frame
\begin{equation} \label{WI01}
u^{\alpha} = \rho (U^{\alpha} + \upsilon^{(i)} e_{(i)}^{\alpha}),
\end{equation}
where $\upsilon^{(i)}$ are the spatial components of the three 
velocity of the fragment
$\rho=1 / \sqrt{1 - |\upsilon |^{2}}$ and $|\upsilon |^{2} = \upsilon^{(i)}\upsilon_{(i)}$. Since spacetime allows the time like 
Killing vector $\xi_{\alpha}=\partial / \partial x^{0}$, it can be 
represented in the tetrad frame as
\begin{equation}\label{WI02}
\xi_{\alpha}=\xi_{(0)}U_{\alpha} + \xi_{(i)}e^{(i)}_{\alpha}.
\end{equation}
Now, the energy $\mathcal{E}$ in terms of the Killing vector can be written as
\begin{equation}\label{WI03}
\mathcal{E} = \xi_{\alpha} U^{\alpha} = \xi_{(0)} = \xi^{\alpha}U_{\alpha} = \xi^{(0)},
\end{equation}
and
\begin{equation}\label{WI04}
g_{00} = \xi^{\alpha} \xi_{\alpha} = -\xi_{(0)}^{2} + \xi_{(i)} \xi^{(i)} = -\mathcal{E}^{2} + |\xi|^{2}.
\end{equation}
Thus, we have the relation
\begin{equation}\label{WI05}
|\xi|^{2} = \xi_{(i)} \xi^{(i)} = \mathcal{E}^2+g_{00}.
\end{equation}
From Eq. (\ref{WI01}), we obtain
\begin{equation}\label{WI06}
\mathcal{\varepsilon} = \xi_{(\alpha)}u^{(\alpha)} = \rho(\xi_{(0)} 
+ \upsilon^{(i)} \xi_{(i)}) = \rho(\mathcal{E} + |\upsilon| |\xi| \cos\theta),
\end{equation}
where $\theta$ is the angle between $\upsilon^{(i)}$ and $\xi_{(i)}$. Using Eq. (\ref{WI05}), one can rewrite Eq. (\ref{WI06}) in the 
following form
\begin{equation}\label{WI07}
\mathcal{\varepsilon} = \rho \mathcal{E} + \rho |\upsilon| \sqrt{\mathcal{E}^{2}+g_{00}}\cos\theta.
\end{equation}
Equation (\ref{WI07}) provides the inequality
\begin{equation}\label{WI08}
\rho \mathcal{E} - \rho |\upsilon| \sqrt{\mathcal{E}^{2}+g_{00}} 
\leq \mathcal{\varepsilon} \leq \rho \mathcal{E} + \rho |\upsilon| 
\sqrt{\mathcal{E}^{2}+g_{00}}.
\end{equation}
For the brane Kerr BH, the Wald inequality can be written in the following form
\begin{equation}\label{WI09}
\rho \mathcal{E} - \rho |\upsilon| \sqrt{\mathcal{E}^{2}+1-b} \leq 
\mathcal{\varepsilon} \leq \rho \mathcal{E} + \rho |\upsilon| \sqrt{\mathcal{E}^{2}+1-b}.
\end{equation}
For $a=\sqrt{2}M$ and $b=-M^2$, radius of the stable innermost 
circular orbit is located at $M$, i.e., $r_{ISCO}\simeq M$ \cite{Aliev}. Substituting the value of $r_{ISCO}$ in Eq. 
(\ref{Ang2}), we can obtain the maximum energy for a particle orbiting a stable circular orbit
\begin{equation}
\mathcal{E}_{0}=\frac{1}{\sqrt{3 - b}}.
\end{equation}
For $\mathcal{\varepsilon}$ to be negative, it is necessary that
\begin{equation}\label{WI11}
|\upsilon| > \frac{ \mathcal{E}}{\sqrt{\mathcal{E}^{2}+1-b}}=\frac{1}{2-b}.
\end{equation}
Otherwise, the fragments should have relativistic energies which 
could be possible before the energy extraction.
For $b=0$, Eq. (\ref{WI11}) reduces to the Kerr BH \cite{Chandrasekhar}.
%%%----------------------------------------------------------%%
\subsection{Efficiency of The Process}
\begin{figure*}
    \begin{minipage}[b]{0.58\textwidth} \hspace{-0.0cm}
        \includegraphics[width=0.8\textwidth]{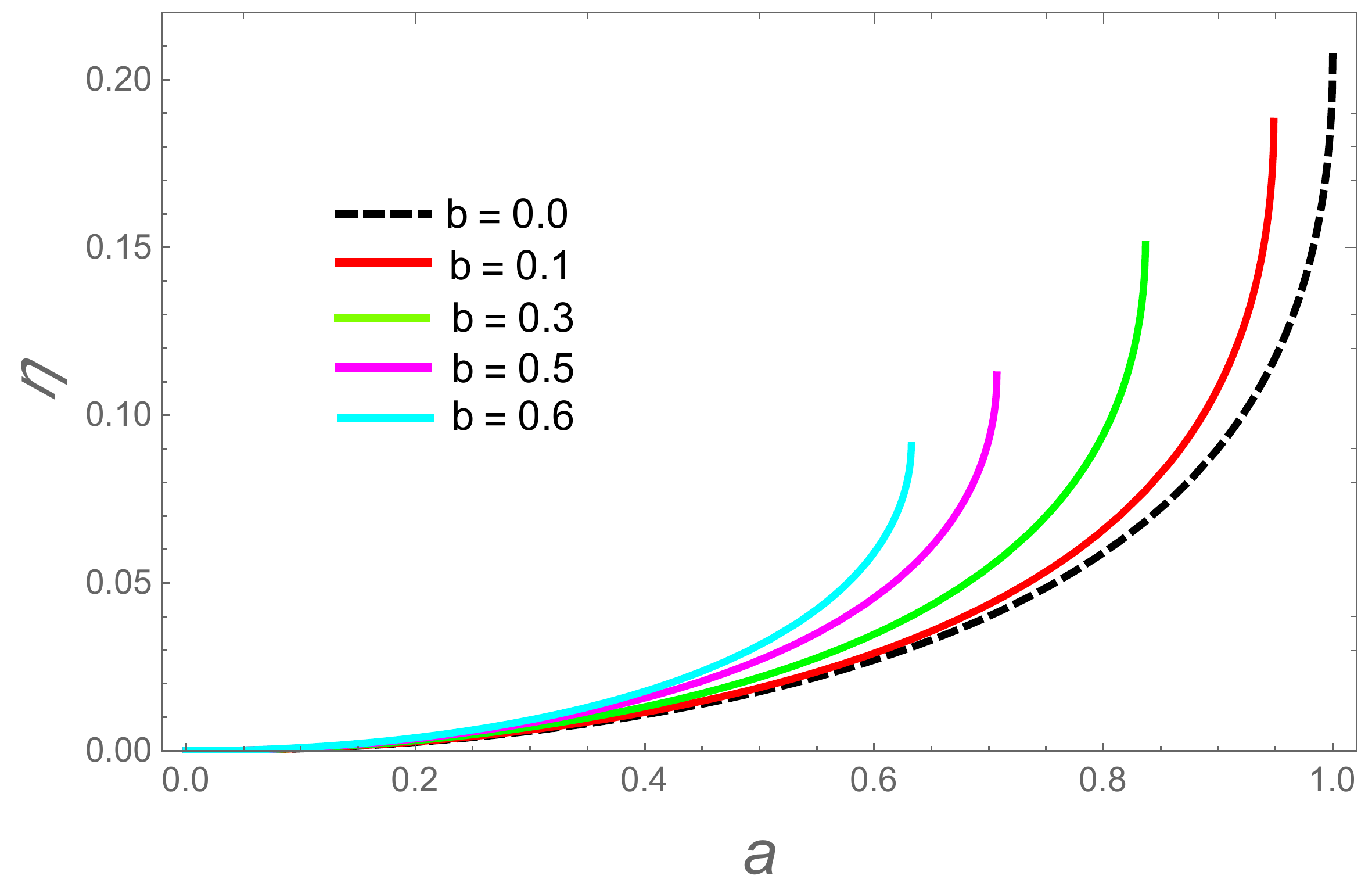}
    \end{minipage}
    \vspace{0.35cm}
        \begin{minipage}[b]{0.58\textwidth} \hspace{-1.5cm}
        \includegraphics[width=.8\textwidth]{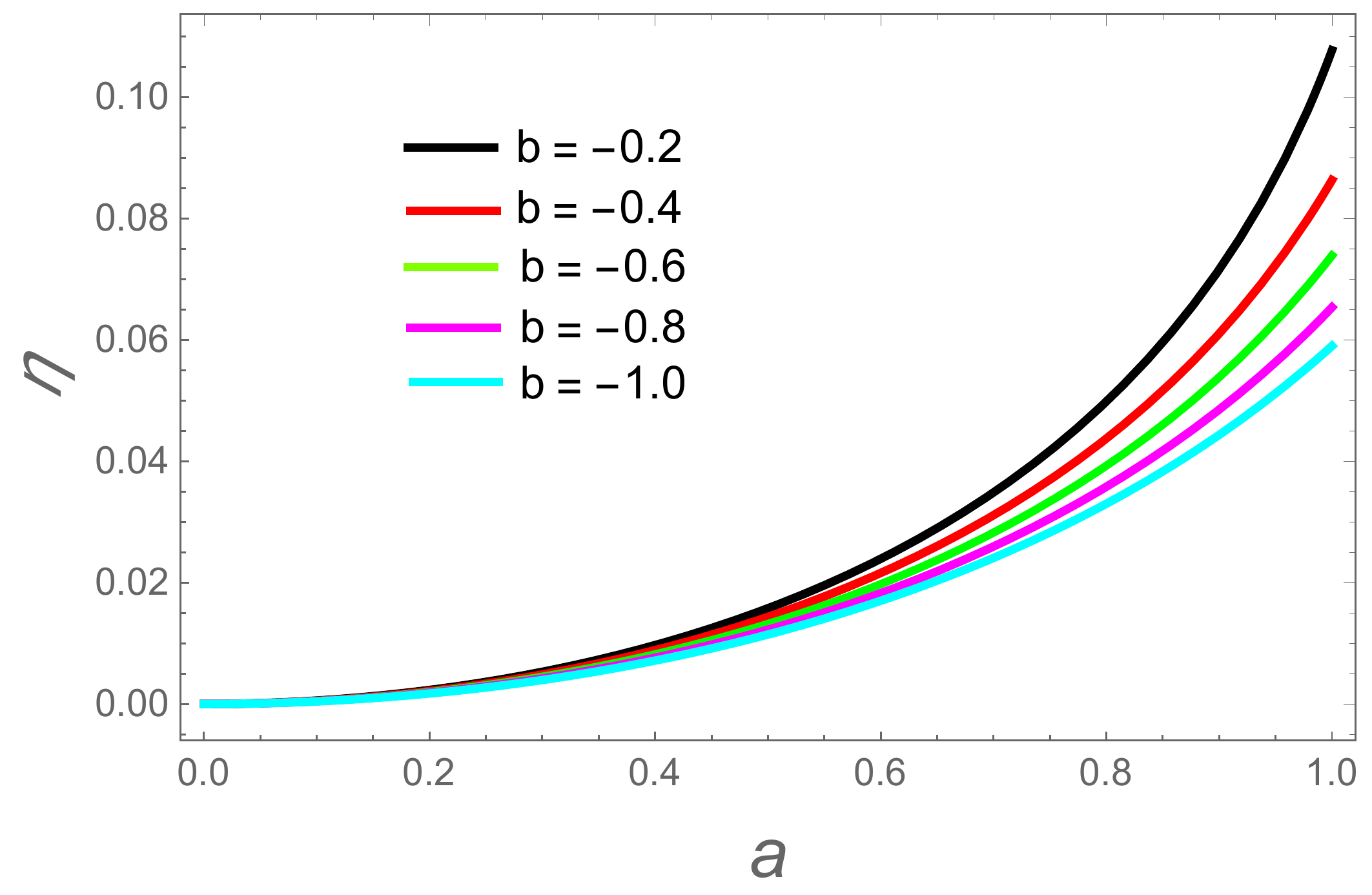}
    \end{minipage}
%%-----------------------------%%
\begin{center}
    \begin{minipage}[b]{0.58\textwidth}
        \includegraphics[width=.8\textwidth]{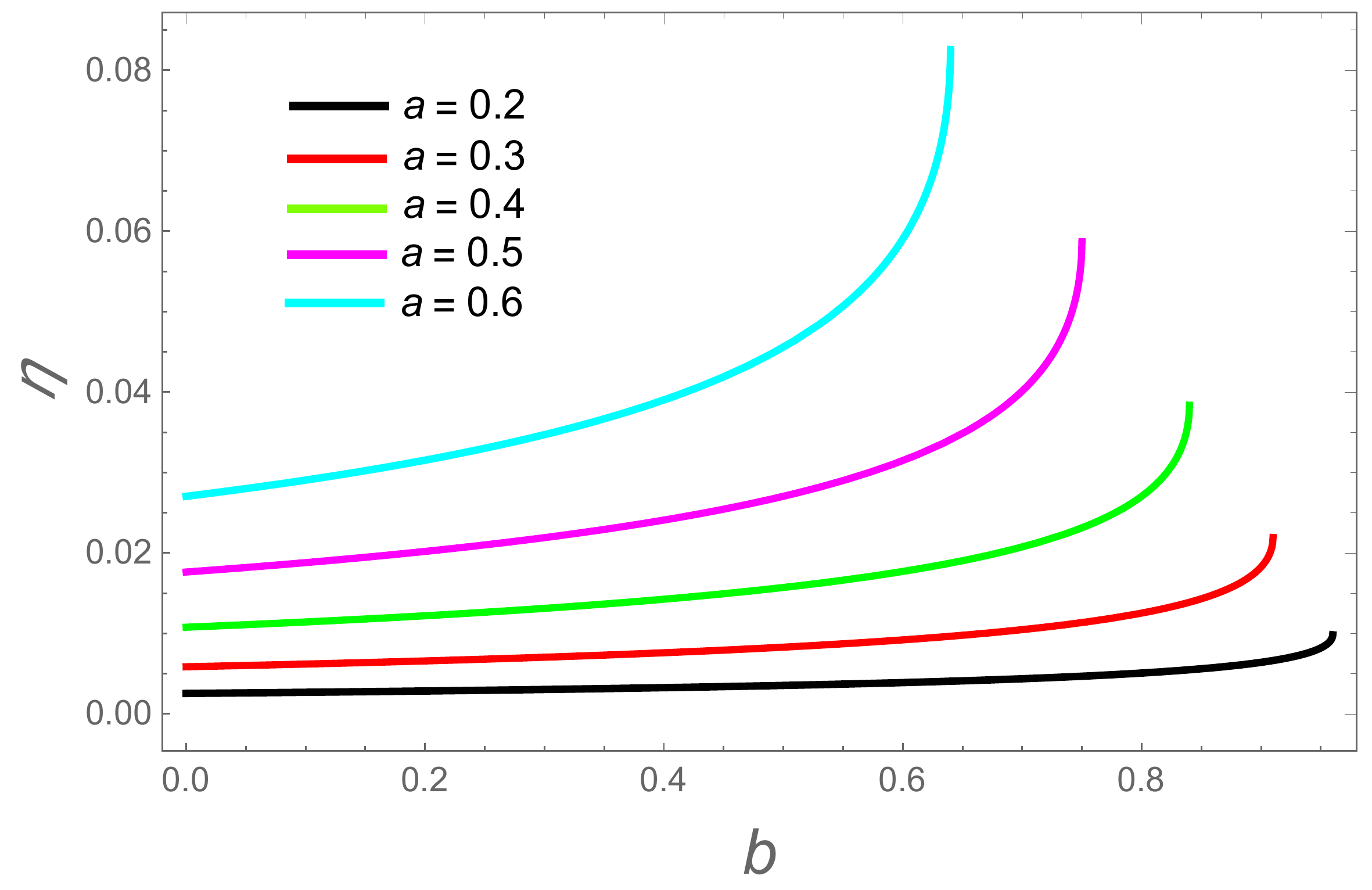}
    \end{minipage}
    \caption{Maximum efficiency of the energy extraction as a function of $a$ (top) whereas $b$ (bottom).}\label{NE02}
\end{center}
\end{figure*}

%%%----------------------------------------------------------%%
The efficiency of energy extraction from a BH via the Penrose 
process is among the principal consequences in the energetics of 
BHs. Consider a particle bearing energy $E^{(0)}$ enters into the 
ergosphere of a BH and subdivide into two particles namely 1 and 2, 
(bearing energy $E^{(1)}$ and $E^{(2)}$, respectively). The particle 
1 has more energy as compared to the incident ones and exits the 
ergosphere whereas, particle 2 bearing negative energy falls into 
the BH. Using the law of energy conservation
\begin{equation} \nonumber
E^{(0)}=E^{(1)}+E^{(2)},
\end{equation}
here $E^{(2)}<0$, implies that $E^{(1)} > E^{(0)}$. Let $\nu=dr/dt$ 
be the radial velocity of a particle with respect to an observer at 
infinity. Using the laws of conservation of angular momentum and energy
\begin{equation}\label{Eff1}
L=p^{t}\Omega, \quad E=-p^{t}Y,
\end{equation}
where
\begin{equation}
Y \equiv g_{tt} + g_{t\phi}\Omega.
\end{equation}
Using $p^{\eta} p_{\eta}=-m^{2}$, we have
\begin{equation}\label{Eff2}
g_{tt}\dot{t}^{2}+2g_{t\phi}\dot{t}\dot{\phi}+g_{rr}\dot{r}^{2}+g_{\phi\phi}\dot{\phi}^{2}=-m^{2}.
\end{equation}
Dividing Eq. (\ref{Eff2}) by $\dot{t}^{2}$, we obtain
\begin{equation}\label{Eff3}
g_{tt}+2 \Omega\ g_{t\phi}+\Omega^{2}g_{\phi\phi} +\frac
{\upsilon^{2}}{\Delta}r^{2}=-\left(\frac{m
Y}{E}\right)^{2}.
\end{equation}
Since the fourth term, in left hand side of Eq. (\ref{Eff3}) is 
always positive and the right hand side is negative or equal to 
zero. Hence, we can rewrite Eq. (\ref{Eff3}) in the following form
\begin{equation}\label{Eff4}
g_{tt}+2 \Omega\ g_{t\phi}+\Omega^{2}g_{\phi\phi}=-\left(\frac{m
Y}{E}\right)^{2}-\frac {\upsilon^{2}}{\Delta}r^{2}\leq 0.
\end{equation}
Using Eq. (\ref{Eff1}), the relation of conservation of angular 
momentum and energy could be expressed as
\begin{eqnarray}\label{Eff5}
p^{t}_{(0)}Y_{(0)}=p^{t}_{(1)}Y_{(1)}+p^{t}_{(2)}Y_{(2)},
\end{eqnarray}
\begin{equation}\label{Eff6}
p^{t}_{(0)}\Omega_{(0)}=p^{t}_{(1)}\Omega_{(1)}+p^{t}_{(2)}\Omega_{(2)}.
\end{equation}
Henceforth, efficiency $(\eta)$ of the collisional Penrose process 
can be described as
\begin{equation}
\eta=\frac{E^{(1)}-E^{(0)}}{E^{(0)}}=\chi-1,
\end{equation}
in which $\chi=E^{(1)}/E^{(0)}$ and $\chi>1$. By the use of Eqs. (\ref{Eff1}),
(\ref{Eff5}) and (\ref{Eff6}), we acquire
\begin{equation}\label{Eff7}
\chi=\frac{E^{(1)}}{E^{(0)}}=\frac{(\Omega_{(0)}-
\Omega_{(2)})Y_{(1)}}
{(\Omega_{(1)}-\Omega_{(2)})Y_{(0)}}.
\end{equation}
Let us consider an incident particle having energy $E^{(0)}=1$ 
enters into the ergosphere and subdivided into two photons having 
momenta $p^{(1)}=p^{(2)}=0$. It can be observed from  Eq. (\ref{Eff7}), that efficiency could be maximized by having the 
smallest value of $\Omega_{(1)}$ and the largest value of $\Omega_{(2)}$ concurrently, which required $\upsilon_{(1)}=
\upsilon_{(2)}=0$. For this case
\begin{equation}\label{Eff8}
\Omega_{(1)}=\Omega_{+}, \quad \Omega_{(2)}=\Omega_{-}.
\end{equation}
The corresponding values of parameter $Y$ are
\begin{equation}\label{Eff9}
Y_{(0)}=g_{tt}+\Omega_{(0)} \ g_{t\phi}, \quad
Y_{(2)}=g_{tt}+\Omega_{-} \ g_{t\phi}.
\end{equation}
The four-momenta of pieces are
\begin{equation}\nonumber
p_{\eta}=p^{t}(1,0,0,\Omega_{\eta}),\quad \eta=1,2.
\end{equation}
Accordingly, Eq. (\ref{Eff2}) could be written as
\begin{eqnarray}
(g_{t\phi}^{2}+g_{\phi\phi})\Omega^{2} + 2\Omega(1+g_{tt})g_{t\phi}+(1+g_{tt})g_{tt}=0.
\end{eqnarray}
Consequently, from the aforementioned equation, the angular velocity 
of the incident particle will take the form
\begin{equation}
\Omega_{(0)}=\frac{-(1+g_{tt})g_{t\phi}+\sqrt{(1+g_{tt})(g^{2}_{t\phi}-g_{\phi\phi}g_{tt})}}{g^{2}_{t\phi}+g_{\phi\phi}}.
\end{equation}
Substituting Eqs. (\ref{Eff8}) and (\ref{Eff9}) into (\ref{Eff7}), 
the efficiency of the energy extraction can be obtained as
\begin{equation}
\eta=\frac{(g_{tt}+g_{t\phi} \Omega_{+})(\Omega_{(0)}-\Omega_{-})}
{(g_{tt}+g_{t\phi}\Omega_{0})(\Omega_{(+)}-\Omega_{-})}-1.
\end{equation}
%%-----------------------------------------------------------%%
\begin{table*} \vspace{-0.0cm}
\begin{center}
\textbf{Table 2:} Maximum efficiency $\eta_{max}$ $(\%)$ of the
energy extraction from BH via the Penrose process.
\scalebox{1}{
\begin{tabular}{c c c c c c c c c }
\hline \hline \noalign{\smallskip\smallskip}
&  {a=0.2}  &  {a=0.4}  &  {a=0.6}  &   {a=0.8}  &  {a=0.9}  &  {a=0.99}  &  {a=1.0}& \\ \noalign{\smallskip}
\hline \noalign{\smallskip\smallskip}
$b=-0.5$    & 0.2046 & 0.8520 & 2.0625 & 4.1361 & 5.7156 & 7.6822  & 7.9470&\\
$b=-0.2$    & 0.2313 & 0.9711 & 2.3930 & 4.9859 & 7.1611 & 10.2955 & 10.7754&\\
$b=-0.1$    & 0.2422 & 1.0208 & 2.5365 & 5.3903 & 7.9267 & 12.0638 & 12.7936&\\
$b=0.0$     & 0.2545 & 1.0774 & 2.7046 & 5.9017 & 9.0098 & 16.1956 & 20.7107&\\
$b=0.005$   & 0.2551 & 1.0805 & 2.7138 & 5.9310 & 9.7713  & 16.6794 & &\\
$b=0.01$    & 0.2558 & 1.0835 & 2.7231 & 5.9608 & 9.1461 & 17.2818 & &\\
$b=0.02$    & 0.2571 & 1.0897 & 2.7419 & 6.0216 & 9.2894 &  & & \\
$b=0.1$     & 0.2685 & 1.1429 & 2.9059 & 6.5846 & 10.8130& & & \\
$b=0.2$     & 0.2846 & 1.2197 & 3.1536 & 7.5876 & & & & \\
$b=0.5$     & 0.3538 & 1.5714 & 4.5583 & & &  & & \\
\hline\hline \noalign{\smallskip}
\end{tabular}}
 \end{center}\label{ep}
\end{table*}
%%-----------------------------------------------------------%%
In order to acquire maximum efficiency ($\eta_{max}$), it is 
necessary for the
incident particle to be subdivided near the horizon of a BH. Henceforth,
the aforementioned equation turns out to be
\begin{equation}
\eta_{max}=\frac{1}{2}\left(\sqrt{\frac{(2Mr_{+}-b)}{r_{+}^{2}}}-1\right).
\end{equation}
The numerically calculated values for maximum efficiency of the 
energy extraction via the Penrose process at different choices of 
the brane parameter $b$ and spin $a$, are shown in Table {\bf{2}}. 
It is described that maximum efficiency of the mechanism could be achieved by increasing the values of the brane parameter $b$. For 
$a=1$ and $b=0$, the maximum efficiency could be $20.7\%$, i.e., the 
limiting value for extreme Kerr BH \cite{Chandrasekhar}. Rotation of 
the BH has a great influence on the motion and particle collision. The rapid rotation of a BH contributes gain in its energy 
extraction. The behavior of the maximum efficiency of energy 
extraction via the Penrose process is also depicted in Fig. {\bf\ref{NE02}} at different values of rotation as well as the tidal 
charge parameter $b$. The dashed curve corresponds to the Kerr BH. It could be observed that for both negative as well as positive 
values of $b$, the efficiency of energy extraction increases with 
brane parameter $b$ of the BH. It is noted that rotation of the BH 
tends to increase the efficiency of energy extraction.
%%---------------------------------------------------------%%
%%---------------------------------------------------------%%
\section{Concluding Remarks}\label{Sec5}
%%---------------------------------------------------------%%
In this article, particle motion and the collisional Penrose process 
in ergoregion of the braneworld Kerr BH has been explored. The 
circular motion of the test particle plays an essential role to understand the accretion disk theory. Geodesics are important to 
study the particle motion and dynamics of galaxies. We have explored the geodesics for particles orbiting the braneworld Kerr BH. The 
properties of event horizon, static limit and ergoregion of the brane Kerr BH are explored in detail. For both positive as well as 
negative tidal charge, the possible situations are discussed. It is worthily to mention that the negative tidal charge could provide a 
mechanism to spin up the BH for which the rotation parameter must be 
greater than the mass of the BH but in the scenario of GR, this situation is not allowed. It is observed that both the spin as well 
as the tidal charge, effects the area of ergoregion. The sign of 
tidal charge also influences the shape of ergoregion and event horizon. For $b>0$, the ergoregion becomes thick, consequently, the 
static limit and horizon decreases. For $b<0$, the boundary of 
ergosphere extends which makes the BH more energetic with respect to 
the extraction of energy.

Energy extraction from the braneworld Kerr BH via the Penrose process has been studied. We have also examined the negative energy 
states, irreducible mass of the BH and limitations of the energy 
extraction by the Wald inequality. It is observed that negative 
energy is allowed only for negative angular momentum. It is found 
that both the negative tidal charge and rotation of a BH tends to 
increase its negative energy while in case of the positive tidal charge, the particles have more negative energy. In case of the Kerr 
BH, particles have more negative energy as compared to the brane Kerr BH with negative tidal charge.

We have explored the efficiency of energy extraction from the brane 
Kerr BH. It is concluded that for both positive as well as negative 
values of the tidal charge, efficiency of the energy extraction increases. Rotation of the BH has great influence on particles 
motion and collision. In case of rapid rotation, particles can take energy from BH rotation as a result, more energy can be extracted. 
Our obtained results are much similar to that of the Kerr BH surrounded by magnetic field as in both situations, the efficiency 
is greatly influenced by the BH rotation \cite{Dadhich}. Moreover, in the case of magnetic Penrose process, the magnetic effects 
increase the efficiency just like in brane Kerr BH does by the brane 
parameter $b$. In comparison with Kerr BH, more energy can be 
extracted in case of brane Kerr BH, but the maximum efficiency limit 
can be obtained for the extreme Kerr BH, i.e., $20.7\%$.

\subsection*{Acknowledgment}
J. L. Ren is very grateful for the financial support from the National Nature Science Foundation of China (11771407).
%%-----------------------------------------------------------%%

\end{document}